\documentclass{optica-article}

\journal{opticajournal} 

\articletype{Research Article}

\usepackage{lineno}

\begin{document}

\title{Pulsed CW laser for long-term spectroscopic measurements at high power in deep-UV}

\author{Nikita Zhadnov,\authormark{1,*} Artem Golovizin,\authormark{1,2} Irene Cortinovis,\authormark{1} Ben Ohayon,\authormark{1,3} Lucas de Sousa Borges,\authormark{1} Gianluca Janka\authormark{1,4} and Paolo Crivelli\authormark{1}}

\address{\authormark{1}Institute for Particle Physics and Astrophysics, ETH, Zurich, 8093, Switzerland

\authormark{2}P.N. Lebedev Physical Institute, Moscow, 119991, Russia

\authormark{3}Technion - Israel Institute of Technology, Haifa, 3200003, Israel

\authormark{4}Laboratory for Muon Spin Spectroscopy, Paul Scherrer Institute, CH-5232 Villigen PSI, Switzerland}

\email{\authormark{*}nzhadnov@phys.ethz.ch} 


\begin{abstract*} 
We present a novel technique for in-vacuum cavity-enhanced UV spectroscopy that allows nearly continuous measurements over several days, minimizing mirror degradation caused by high-power UV radiation. Our method relies on pulsing of the cavity's internal power, which increases the UV intensity to maximum only for short periods when the studied atom is within the cavity mode volume while keeping the average power low to prevent mirror degradation. Additionally, this method significantly decreases laser-induced background on charged particle detectors. The described 244 nm laser system is designed for 1S-2S two-photon CW spectroscopy of muonium in the Mu-MASS project. It was tested to provide intracavity powers above 20 W, requiring maintenance only a few times a day. The pulsing technique demonstrates minimal impact on the radiation frequency, with no observed shifts exceeding 15 kHz. Our approach represents a promising new technique for high-precision spectroscopy of atoms in harsh UV environments and demonstrates the feasibility of CW spectroscopy of muonium.

\end{abstract*}

\section{Introduction}

High intensity deep-ultraviolet (deep-UV) continuous wave (CW) lasers open up new opportunities for light-matter interaction for testing fundamental physical theories and developments in applied science \cite{Zhang2022KeyLasing,Savage2007UltravioletLasers}. Outperforming pulsed lasers in frequency stability, they are an essential tool for precision spectroscopy of transitions from the ground state in hydrogen \cite{Parthey2011ImprovedFrequency,Fleurbaey2018NewPuzzle,Brandt2022MeasurementHydrogen} and its isotopes \cite{Parthey2010PrecisionShift}, antihydrogen \cite{Ahmadi2018CharacterizationAntihydrogen} and muonium \cite{Crivelli2018TheExperiment,Ohayon2021CurrentPSI,Burkley2021StableCoatings}. 
Deep-UV CW lasers allow laser cooling of hydrogen \cite{Cooper2018Cavity-enhancedHydrogen}, mercury ions and atoms \cite{Liu2019TrappedPhotoionization,Hachisu2008TrappingClocks,Zhang2021AAtoms}, cadmium \cite{Kaneda2016Continuous-waveAtoms}, AlF \cite{Truppe2019SpectroscopicTrapping}, AlCl \cite{Shaw2021StableMonochloride} and HgF \cite{Yang2019Laser-cooledElectron}.
Progress of UV laser technologies is beneficial for development of new optical atomic clocks \cite{Kaneda2016Continuous-waveAtoms,Hachisu2008TrappingClocks,Petersen2008Doppler-freeMercury,Seiferle2019EnergyTransition} and single photon excitation of Rydberg states \cite{Zeiher2016Many-bodyLattice,Wang2017Single-photonCell}.
Collinear resonance ionization spectroscopy using CW lasers \cite{Voss2013FirstProperties,DeGroote2015UseSpectroscopy} is at the forefront of measuring nuclear properties, such as shape, spin, and moments, through hyperfine structure analysis.
Studying certain atomic species with this technique requires the use of deep-UV laser light for their excitation \cite{Neugart2017CollinearHighlights,Gustafsson2020TinStudies}.

Currently, the highest power deep-UV laser systems are capable of generating up to 2 W of CW light \cite{Shaw2021StableMonochloride,Burkley2019HighlySystem}. Moreover, a number of experiments such as the laser cooling of hydrogen \cite{Cooper2018Cavity-enhancedHydrogen} and muonium spectroscopy \cite{Burkley2021StableCoatings} require tens of watts of such radiation, which can be obtained by amplification in an optical cavity. A big challenge that appears at such intensities of UV light, especially in vacuum, is the degradation of optical components: mirrors, crystals, lenses, and windows \cite{Kunz2000ExperimentationOptics,Hollenshead2006ModelingOptics,Gangloff2015PreventingCoatings}.

Cavity mirrors exposed to maximum power suffer the most from degradation. The speed of degradation is known to increase with the UV light power which optical surfaces experience. Previous research \cite{Burkley2021StableCoatings} demonstrated that an ultra-high vacuum (UHV, $10^{-8}$ mbar) Fabry-P\'erot cavity, with oxide multi-layer coatings mirrors ($\mathrm{HfO_{2}/Al_{2}O_{3}}$) on a $\mathrm{SiO_{2}}$ substrate, experiences more than a two-fold decrease of finesse in one hour at $5\,$W of intracavity power. Fluoride-coated mirrors ($\mathrm{MgF_{2}/LaF_{3}}$) on $\mathrm{CaF_{2}}$ substrates demonstrate a much more stable behavior: having slightly worse optical performance, they can maintain up to $10\,$W of laser power in UHV on one-hour timescales. Nevertheless, more than these record-breaking characteristics are required in some applications. One possible way to maintain the optical quality of the mirrors is to operate them in an oxygen atmosphere. This approach demonstrated continuous operation for more than 4 hours at the power of $16\,$W at $10^{-3}\,$mbar of oxygen \cite{Burkley2021StableCoatings}. However, most precision spectroscopy applications demand UHV conditions, which are extremely hard to combine with this technique.
A compromise solution may be periodic conditioning of the mirrors with oxygen in the presence of UV light. This requires breaks in the experiment but allow for restoring optical characteristics between measurement periods. The most likely cause of the degradation of fluoride mirrors is hydrocarbon contamination. During recovery, UV light generates ozone and atomic oxygen and decomposes hydrocarbon contaminants into components. The latter reacts with oxygen atoms and form simpler volatile molecules desorbed from the surface. Unfortunately, working with more than $10\,$W of intracavity power escalates degradation and requires spending more time on mirror recovery than on actual measurements.

Another important issue relates to resonance ionisation spectroscopy \cite{Crivelli2018TheExperiment}. In such experiments, charged particles (ions, muons) are most often detected using microchannel plates (MCP). These devices are sensitive to deep-UV photons. Since it is very problematic to provide complete protection of an MCP from scattered light without reduction of its sensitive area, this effect inevitably leads to a decrease in signal-to-noise ratio.

The laser system presented in this article was developed for the Mu-MASS project aiming at high-precision CW laser spectroscopy of 1S-2S transition in muonium.
In this experiment,  muons, coming from the low energy muons (LEM) beamline at PSI \cite{Prokscha2004APSI}, hit a mesoporous thin film silica target. Around 60 $\%$ are converted into muonium atoms and 40 $\%$ diffuse back into vacuum before decaying\cite{Antognini2012MuoniumTemperatures}. Some of those atoms pass through a standing wave of $244\,$nm light created with an enhancement cavity and can be excited to the 2S state. The 2S atoms are then photoionised with a 355 nm pulsed laser and the resulting muon is detected by an MCP (for details see \cite{Cortinovis2023UpdateFrequency}). Due to this transition's low two-photon excitation cross-section value and the small number of available muonium atoms, having an excitation rate of 1 event per hour requires $25\,$W of laser power on resonance. Therefore, to collect a statistically significant dataset, the high-power laser has to be stable for several days.
The possibility of continuous conditioning of the mirrors for the Mu-MASS project is limited by the fact that oxygen contamination induces degradation of the LEM moderator \cite{Morenzoni2003LowRegions}. Even though differential pumping was implemented between the cavity mirrors and the moderator zone, a reduction of the muon flux by a factor of two was measured in a timescale of a few minutes. Overcoming this obstacle would require impractical and complicated additions to the Mu-MASS vacuum system, involving multiple stages of differential pumping. 
This article introduces a new approach to enable long-term, high-power laser spectroscopic measurements in the deep-UV region for Mu-MASS and similar experiments requiring clean UHV conditions.

\section{Methods}
\begin{figure}[htbp]
\centering\includegraphics[width=13cm]{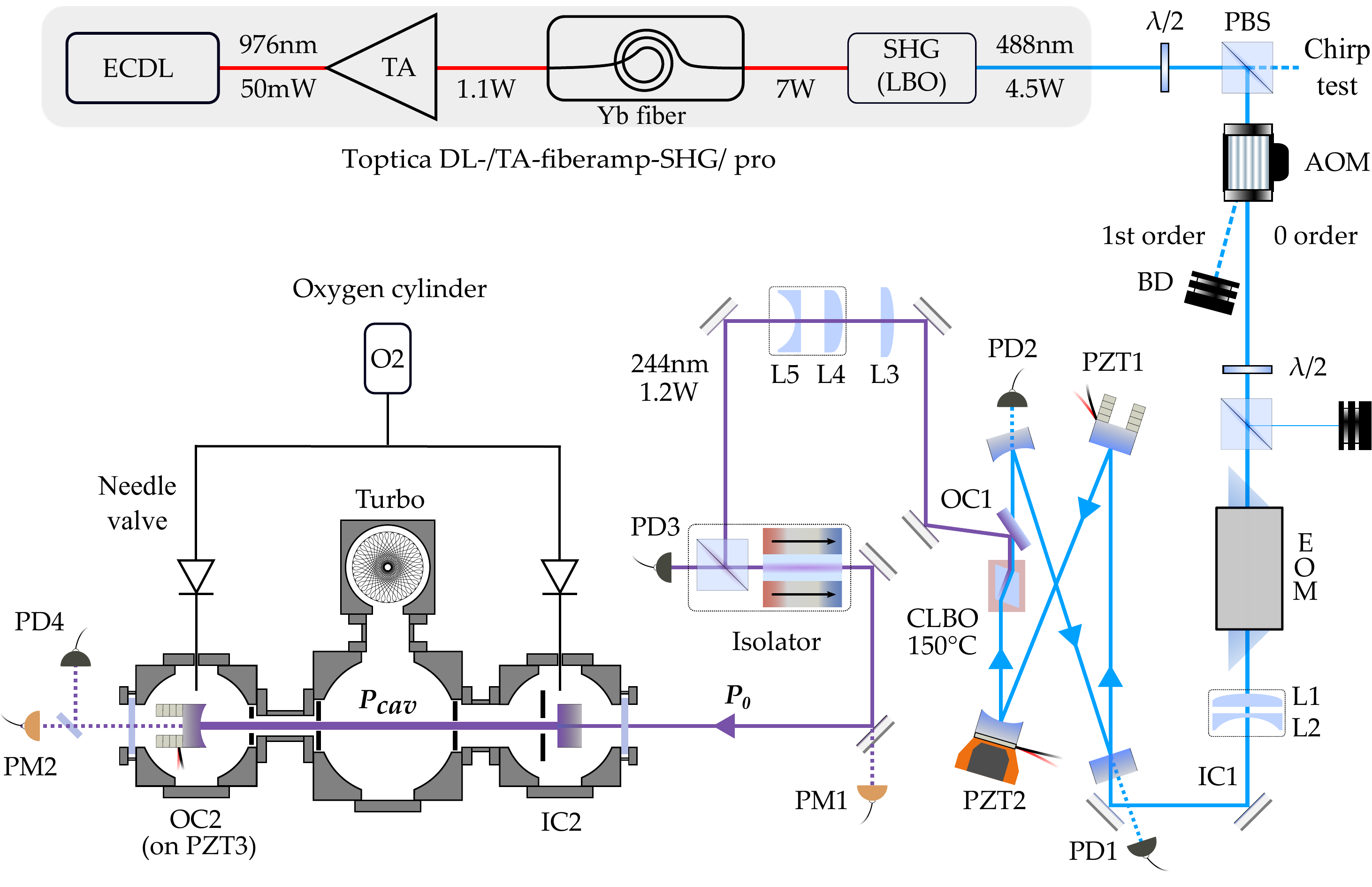}
\caption{Experimental setup for muonium spectroscopy. ECDL - external cavity diode laser, TA - tapered amplifier, SHG - second harmonic generator, LBO - lithium triborate, PBS - polarising beam splitter, AOM - acousto-optic modulator, BD - beam dump, EOM - electro-optic modulator, L1-L5 - mode matching lenses, PD - photodetector, PM - power monitor, CLBO - caesium lithium borate, IC - input coupler, OC - output coupler, HR - high reflector, PZT - piezoelectric transducer stack. Toptica laser assembly generates 4.5 W of $488\,$nm light, which goes through AOM and then converted to 244 nm using a SHG cavity with a CLBO crystal. The resulting UV light, reaching up to 1.2 W, is coupled to an enhancement cavity inside the ultrahigh vacuum chamber.}
\label{fig:setup}
\end{figure}

The experiment was conducted using the Mu-MASS laser setup, as shown on Fig. \ref{fig:setup}. The setup consists of several key components, including a high-power infrared Yb fiber amplifier, a CLBO crystal-based UV second harmonic generator, and a vacuum enhancement cavity. These components have been previously described in \cite{Burkley2017YbNm,Cooper2018Cavity-enhancedHydrogen,Burkley2019HighlySystem,Burkley2021StableCoatings}. For this experiment, fluoride mirrors were chosen for both the input and output couplers of the enhancement cavity due to their durability in UV. To the best of our knowledge, this laser system represents the most powerful $244\,$nm CW laser source currently available.

Spectroscopic measurements often work in repetitive mode: the atoms or molecules of interest are prepared, interact with laser radiation and one measures their excitation efficiency. Usually, it requires the laser light to be available on demand or periodically \cite{Parthey2011ImprovedFrequency,Crivelli2018TheExperiment,DeGroote2015UseSpectroscopy}. Therefore, to slow down the effect of UV-induced degradation without sacrificing the ultimate laser power one can deliver the light only for the necessary measurement period, avoiding constant irradiation of optical components.
A similar technique was used for visible laser for the CRIS (collinear resonance ionization spectroscopy) project \cite{DeGroote2015UseSpectroscopy}, where the laser power was pulsed with a Pockels cell. As CRIS works with a bunched beam, the pulsing of CW light is used to obtain higher spectroscopic resolution while suppressing the background.
The use of cavities for amplification and second harmonic generation creates challenges for the pulsing technique: it does not allow decreasing the laser power below the level necessary to maintain the cavities' length in resonance with the laser.

A laser power control system was implemented using a quartz crystal-based AOM (Fig. \ref{fig:setup}) that is designed for high light intensities. Pulsing the AOM RF amplitude lets us quickly adjust the zero-order beam intensity at 488 nm, with a base-to-peak power level ratio of more than 10. By aligning the AOM to maximize the first diffraction order (with a diffraction efficiency of over $>90\%$), we direct the zero order to the UV SHG and stop the first order with a beam dump.
Bragg diffraction of light on a sound wave leads to spoilage of the zero-order mode shape by ``eating away" the intensity in the central part of the beam more than at the edges. This effect is advantageous in reducing light coupling to the SHG not only by direct power reduction but also by worse mode matching. Using this technique and taking into account the nonlinear efficiency of second harmonic generation, we are able to maintain a baseline power of UV of just a few mW, which can peak up to $1.2\,$W when the AOM is not active. The optimal pulse duration was determined by computer simulation of muonium formation and its escape from the target to maximize the probability of 1S-2S excitation. The resulting value is around 1 microsecond. To keep the SHG and the enhancement cavity in resonance with the laser wavelength, Pound-Drever-Hall locking scheme \cite{Drever1983LaserResonator} is used. The bandwidths of the feedback loops are constrained by the first resonance frequencies of the piezoelectric transducers utilized to regulate the positions of the mirrors. The enhancement cavity uses a center-of-mass piezo transducer mount \cite{Chadi2013Note:Components}, which provides the widest feedback bandwidth, however still less than $100\,$kHz. As these cavity-locking systems are slow to react to $1\,\mu$s-scale pulsing of laser power, they act as low-pass filters for such fast disturbances, allowing the cavities to remain locked throughout the measurement. In the described laser system, even few-tens-of-microseconds-long pulses do not cause the cavities to unlock.

\begin{figure}[htbp]
\centering\includegraphics[width=8cm]{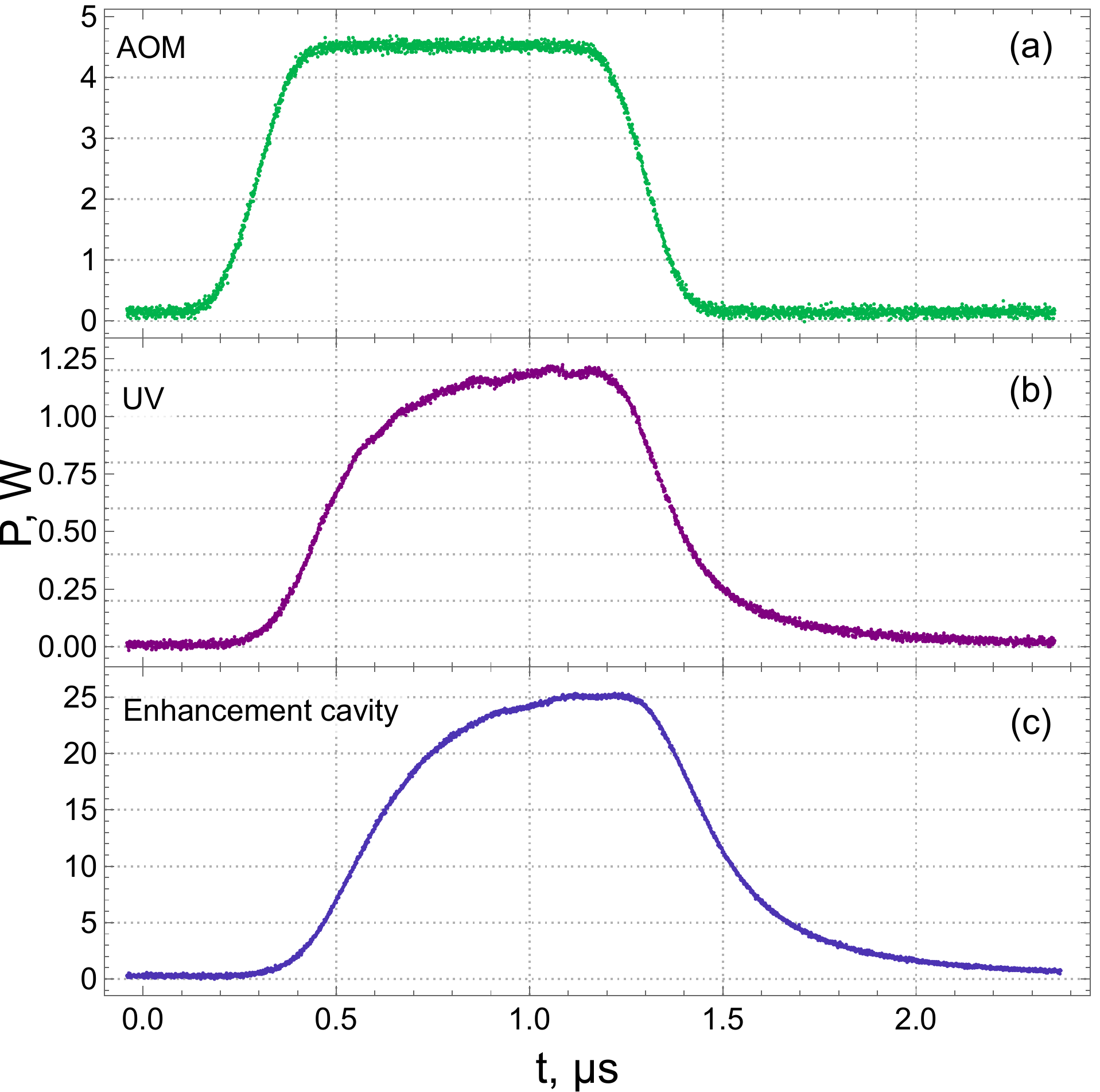}
\caption{The laser power pulse is measured at three positions in the optical scheme: (a) in the zero-order right after the AOM, (b) incoming to the enhancement cavity, and (c) after transmission through the enhancement cavity. The RF pulse, which triggers the AOM, occurs at $t=0$ s, and the delay of the laser pulse is caused by the travel time of the sound wave in the AOM crystal and the exponential behavior of power increase after passing through the SHG cavity and the enhancement cavity, with finesses of 150 and 200 (non-degraded), respectively. Eventually, the intracavity UV power response is characterized by a delay of 460 ns, followed by an exponential rise with a time constant of 190 ns.}
\label{fig:pulses}
\end{figure}

The accurate control of delays in the optical and radio wave lines is crucial for the successful implementation of the described method. Figure \ref{fig:pulses} illustrates a single laser power pulse at three different points in the optical setup: at the zero-order of the AOM, after the UV SHG, and inside the enhancement cavity (see Figure \ref{fig:setup}). Intracavity power was determined from the transmitted light power measurement, taking into account the transmission coefficient of the output coupler. To minimize the AOM delay, we positioned the laser beam as close as possible to the ultrasonic emitter on the edge of the AOM crystal, right up to the point where the beam starts to cut into the crystal. Due to the delay and the time constants of the cavities, it takes $1\,\mu$s for the UV light to reach 95\% of its peak power level after the AOM RF amplitude pulse. Much shorter delays, rise times and higher extinction ratios can be achieved with the use of Pockels cell instead of AOM \cite{DeGroote2015UseSpectroscopy}. In that case, the power rise time would be limited by the cavities.

A triggering rate of 5 kHz (typical for the LEM muon beam at PSI), results in an average duty cycle of $5\times10^{-3}$ for $1\,\mu$s long laser pulses.
With a base UV power of $5\,$mW and a peak power of $1.2\,$W, this enables the average power to be reduced by more than 100 times compared to the peak power. 
In UHV conditions, even this huge reduction in the UV load on the mirrors does not allow overcoming the degradation effect completely, but should dramatically reduce its rate. 
A second interesting feature, which should be provided by fast AOM power control, is the opportunity to suppress photon-induced background signals on MCP detectors.

\section{Results and discussion}

The performance of the laser system was monitored during the first attempt of muonium CW spectroscopy. The lasers and the vacuum chamber were transported from ETH Zurich to PSI. The task for the laser system during the measurements was to work stably at high peak power for several days.

\begin{figure}[htbp]
\centering\includegraphics[width=10cm]{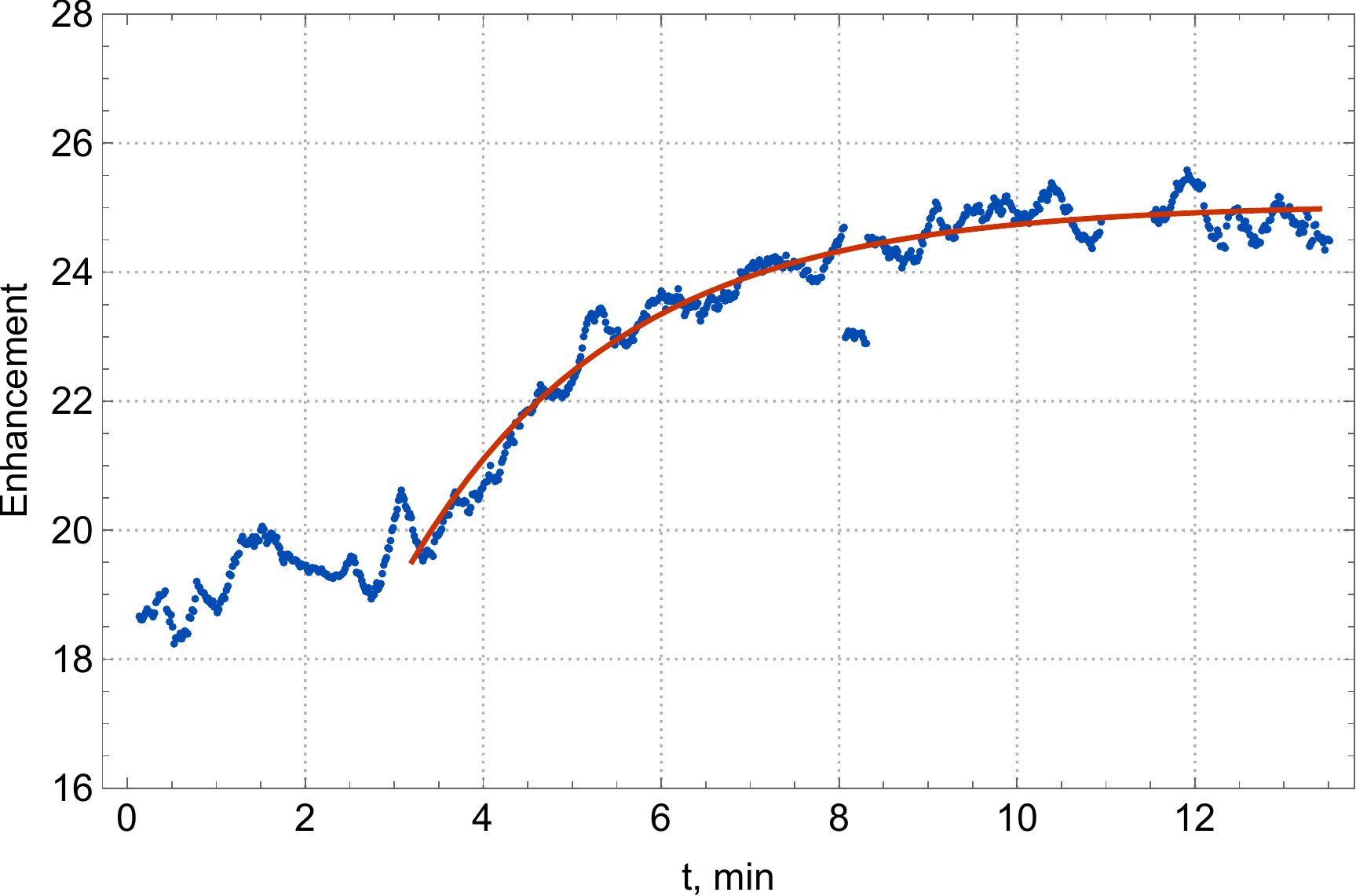}
\caption{The cavity's enhancement factor during oxygen conditioning was determined by measuring the cavity transmission. The conditioning was performed with $1\,$W of intracavity UV light and an oxygen pressure of $10^{-2}\,$mbar. The recovery process had a time constant of $2.5\,$minutes. The fluctuations on the graph are due to the influence of the oxygen flow from the needle valves.}
\label{fig:conditioning}
\end{figure}

To maintain a high enough enhancement factor in the course of UV laser light operation, the cavity mirrors need recovery from hydrocarbon contamination \cite{Burkley2021StableCoatings}. During the recovery process, the vacuum chamber of LEM was shut off to protect the moderator, and oxygen was pumped in through the needle valves near the mirrors (Fig. \ref{fig:setup}) up to the pressure of $10^{-2}\,$mbar. At an intracavity power of about $1\,$W the enhancement recovery process has a time constant of 2.5 minutes (Fig. \ref{fig:conditioning}). The entire procedure including oxygen evacuation down to $10^{-6}\,$mbar was taking almost half an hour, which is fast enough to not significantly affect the duration of the measurements.

\begin{figure}[htbp]
\centering\includegraphics[width=10cm]{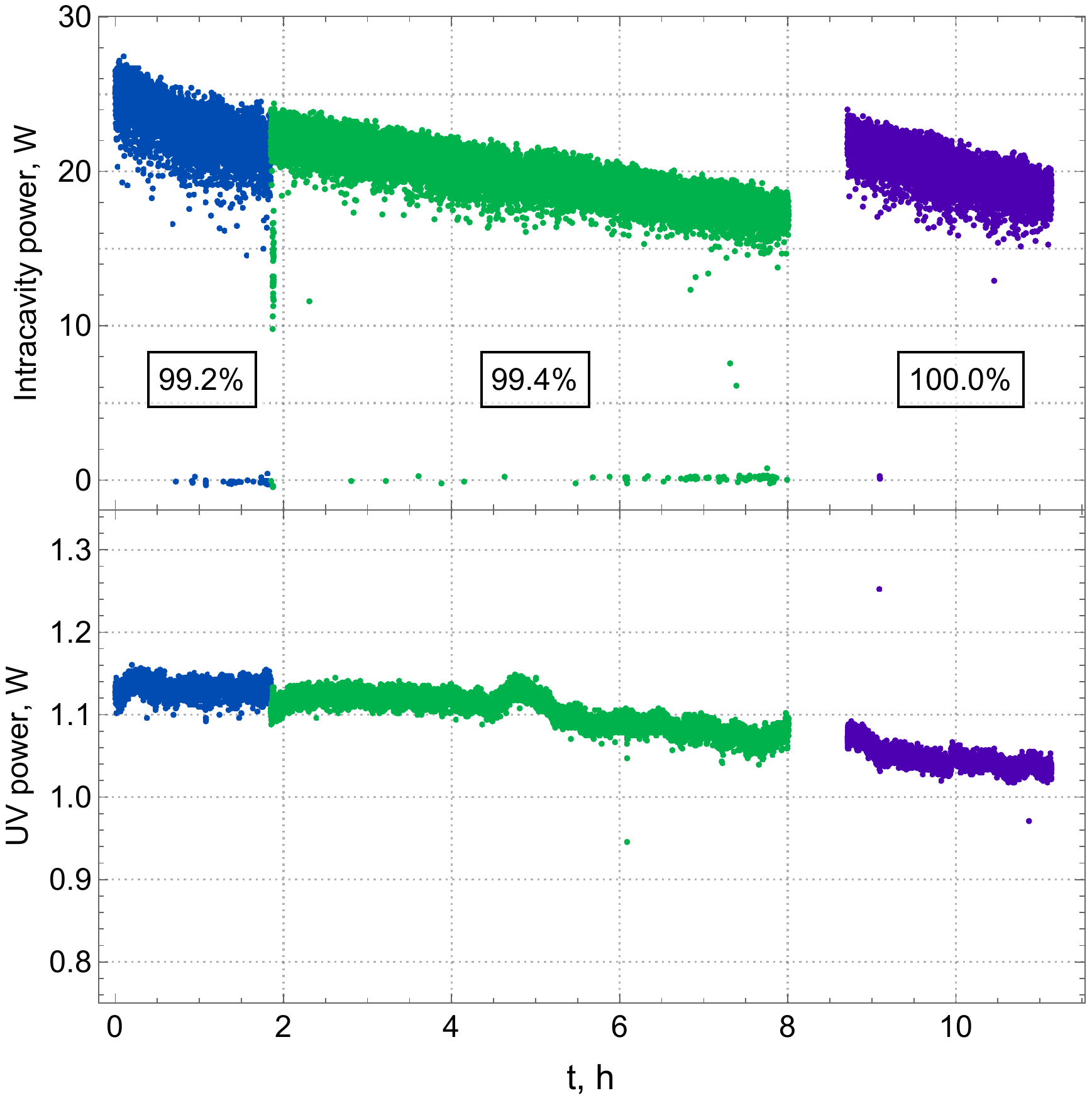}
\caption{Fragment of intracavity and UV SHG output peak power measurements. At $t=2\,$h the quality of the enhancement cavity lock was improved, and at $t=8$ h mirror recovery started. In the first graph, the percentage of uptime is shown. Fluctuations of the peak power are caused by imperfect cavity lock quality and vibrations. The descending intracavity power is due to mirror degradation and a decrease of UV generation efficiency.}
\label{fig:timeline 11h}
\end{figure}

The UV SHG output radiation was operated at the base power of $5\,$mW, responding to each trigger signal with a pulsed increase in power up to more than $1\,$W. The average trigger rate during the experiment was $2.5\,$kHz. The incoming and transmitted laser peak power were logged during the whole measurement to monitor the real-time cavity enhancement. A fragment of these measurements is presented in Fig. \ref{fig:timeline 11h}. The experiment showed that with an initial intracavity pulse power of 25 watts, the resonator degrades to 20 watts in approximately 5 hours and requires recovery (Fig. \ref{fig:timeline 11h}). A notable observation is that there were no signs of degradation on the output coupler of the UV SHG during the pulsing regime, which had been a concern in the past.

We should mention that the enhancement factor and UV SHG output power in \cite{Burkley2021StableCoatings} is larger than in this work. During the measurement, the enhancement was worsened by vibrations that disturb the coupling of laser light into the cavity mode. The piezo's ability to suppress vibrations will be improved by shortening the cavity and placing it vertically. The second harmonic generator suffered from dust contamination and misalignment due to temperature changes, showing slightly worse behavior than optimal. Dust protection and periodic alignment help to maintain its stable operation.

The laser system was operating for about 80 hours during five days of measurements. Breaks in work were mainly due to interruptions in the operation of the muon beam, background noise tests, and other measurements in which the laser was not involved. It is worth noting that the installation of the laser setup in the LEM beamline took around 40 hours, and even during the days of measurement, misalignments due to laser transportation were improved over the course of the measurement.

During propagation through a transparent medium, a laser pulse can acquire a chirp due to the effects of chromatic dispersion, nonlinearities, or rapid, laser-induced changes of the refractive index. 
To analyze this effect, we assembled an additional setup for UV light frequency measurement. 
This includes an alternative beam path where the laser light does not experience pulsing. 
One watt of blue light produced by the Toptica box is split off with a polarizing beamsplitter (see Fig. \ref{fig:setup} after LBO SHG) and directed to another AOM, which provides detuning, and then to a single-pass BBO crystal to generate 244 nm light. 
The two beams are combined and directed to a photodiode, which produces a beat signal at a doubled detuning frequency. 
In our experiment, this was $124.445\pm0.002\,$MHz. 
Since we are interested in frequency changes smaller than the muonium 1S-2S transition linewidth (144 kHz), a single laser pulse waveform does not provide enough data due to the spectral limit set by its short duration.
Furthermore, the silicon photodetector used in the deep-UV region has a very moderate sensitivity, and the second harmonic produced by the BBO crystal has low intensity, which reduced the signal-to-noise ratio. 
To collect a statistically significant sample, we analyzed 2000 waveforms. 
We took the first $750\,$ns of each pulse, filtered them with a band-pass Fourier filter, and calculated the average frequency by measuring the difference between neighboring zero-crossings of the sine signal at the beat frequency. This gave us a normally-distributed sample of two thousand instantaneous frequencies, which were used to obtain an average value of $124.445\pm0.006\,$MHz. This allows us to conclude with a 95\% confidence level that no frequency shift above $15\,$kHz is introduced by the pulsing scheme.

\begin{figure}[htbp]
\centering\includegraphics[width=8cm]{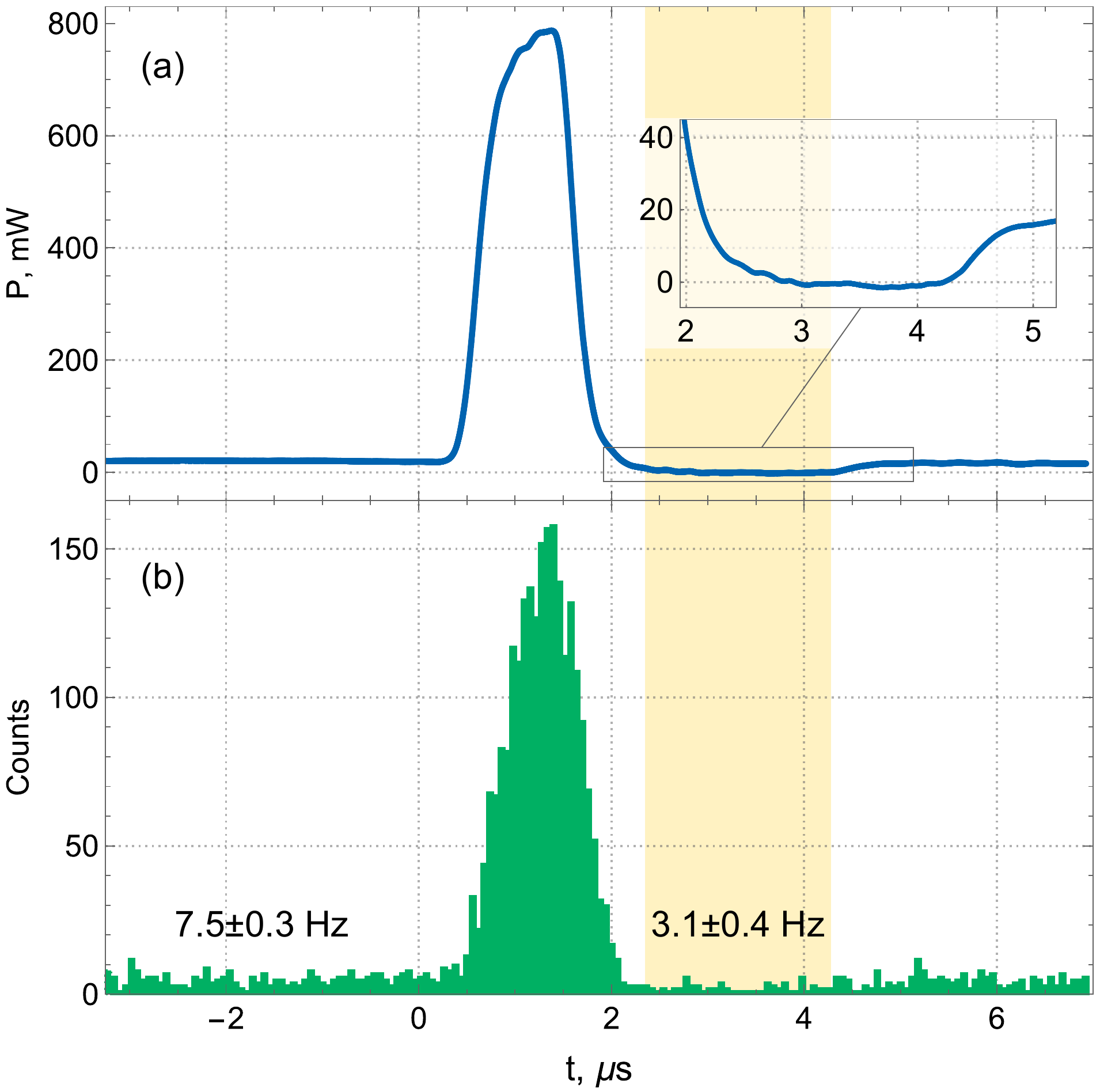}
\caption{(a) Power pulse inside the enhancement cavity followed by a dip. (b) The temporal distribution of the number of counts detected by the MCP detector. The region where the radiation power is minimal is highlighted. For the description of the Mu-MASS experiment see \cite{Cortinovis2023UpdateFrequency}.}
\label{fig:background}
\end{figure}

The control of AOM power not only enables pulsing of UV intensity but also allows for an almost complete turn-off of the UV light by maximizing AOM diffraction efficiency (Fig. \ref{fig:background}(a)). This means that the UV light intensity inside the vacuum chamber can be decreased by at least a factor of 10 in comparison to the base level for a few microseconds after the pulse, creating a reduced background time window for less noisy particle detection. 
To reach the MCP detector, which is removed at a distance of half a meter from the enhancement cavity, photons need to pass through a 90-degree bend in the vacuum tube \cite{Cortinovis2023UpdateFrequency}. 
For extra protection of the detectors from scattered laser radiation, we installed four light-absorbing screens (Acktar MaxiBlack foil) with axial holes in the vacuum chamber. 
These screens isolate the mirror area from the central chamber (Fig. \ref{fig:setup}).
The signal of the MCP was measured for 12 hours with $1\,\mu$s intracavity UV peak power of $800\,$mW and a following $3\,\mu$s long minimum-power time window. Turning the radiation off made it possible to reduce the background rate by a factor of two, as shown in Figure \ref{fig:background}(b): from $7.5\pm0.3\,$Hz to $3.1\pm0.4\,$Hz. 

\section{Conclusion}

In summary, this study introduces a new and innovative method for precision continuous wave laser ultraviolet spectroscopy that addresses the challenge of optical surface degradation commonly encountered in such experiments. On-demand CW UV laser light pulsing enables the use of cavity-enhanced radiation in ultrahigh vacuum conditions, with a power output range of $20-25\,$W. Spectroscopic measurements with this technique can last at least a week. To maintain a high enhancement coefficient, mirror recovery with oxygen conditioning is only required 4-5 times a day, which has minimal impact on the measurement process. Precise control over UV radiation power is beneficial for resonance ionization spectroscopy, as it allows for rapid turn-offs of the laser radiation to minimize UV pickup by MCP detectors during the registration time window.

The presented laser system was designed for 1S-2S spectroscopy of muonium in the Mu-MASS project. With minor improvements, the pulsed CW laser system demonstrates the feasibility of muonium continuous-wave spectroscopy and paves the way for further studies of this unique atomic system.

\begin{backmatter}
\bmsection{Funding}
This work is supported by the ERC consolidator grant 818053-Mu-MASS (P.C.) and the Swiss National Science Foundation under grant 197346 (PC). B.O. acknowledges support from the European Union’s Horizon 2020 research and innovation program under the Marie Skłodowska-Curie grant agreement No. 101019414. 
\bmsection{Acknowledgments}
We would like to acknowledge Zak Burkley for his essential contribution of the laser system. We would like to thank Dylan Yost for the very useful discussions. 

\bmsection{Disclosures}
The authors declare no conflicts of interest.

\bmsection{Data Availability Statement}
Data underlying the results presented in this paper are not publicly available at this time but may be obtained from the authors upon reasonable request.
\bigskip
\end{backmatter}


\bibliography{references}






\end{document}